\documentclass[letterpaper]{aa}
\usepackage{natbib}
\usepackage{graphicx}
\usepackage{rotating}
\bibpunct{(}{)}{;}{a}{}{,}
\begin{document}
\title{Gas-grain chemistry in cold interstellar cloud cores with a microscopic Monte Carlo approach to surface chemistry}
\author{Q. Chang\inst{1}, H. M. Cuppen\inst{1,2}, and E. Herbst\inst{1,3}}
\institute{Department of Physics, The Ohio State University,
Columbus, Ohio 43210, USA
           \and Leiden Observatory, Leiden University, P. O. Box 9513, 2300 RA Leiden, The Netherlands                  
           \and Departments of
Chemistry and Astronomy, The Ohio State
          University,
Columbus, OH 43210, USA} 

\offprints{herbst@mps.ohio-state.edu}
         \date{Received
{\today}}

  \abstract%
 {}
{We have recently developed a microscopic Monte Carlo approach  to study surface chemistry on interstellar grains and the morphology of  ice mantles.  The method is  designed to eliminate the problems inherent in the rate-equation formalism to surface chemistry. Here we report the first use of this method in a chemical model of cold interstellar cloud cores that includes both gas-phase and surface chemistry. The surface chemical network consists of a small number of diffusive reactions that can produce molecular oxygen, water, carbon dioxide, formaldehyde, methanol and assorted radicals.}
 {The simulation is started by running a gas-phase model including accretion onto grains but no surface chemistry or evaporation.  The starting surface consists of either flat or rough  olivine.   We introduce the surface chemistry of the three species H, O and CO in an iterative manner using our stochastic technique.   Under the conditions of the simulation, only atomic hydrogen can evaporate to a significant extent.   Although it has little effect on other gas-phase species, the evaporation of atomic hydrogen changes its gas-phase abundance, which in turn changes the flux of atomic hydrogen onto grains.  The effect on the surface chemistry is treated until convergence occurs. We neglect all non-thermal desorptive processes.}
 {We determine the mantle abundances of assorted molecules as a function of time through $2 \times 10^{5}$ yr.  Our method also allows determination of the abundance of each molecule in specific monolayers.  The mantle results can be compared with observations of water, carbon dioxide, carbon monoxide, and methanol ices in the sources W33A and Elias 16.  Other than a slight underproduction of mantle CO, our results are in very good agreement with observations.
}
 {}

\titlerunning{Stochastic gas-grain chemistry}
\authorrunning{Chang et al.}


\maketitle

\section{Introduction}
Not only are interstellar grain surfaces believed to be the sites  where molecular hydrogen is 
formed in diffuse clouds, but in cold cores of dense clouds interstellar grain chemistry is an 
important mechanism for the formation of quite a few additional species, most of which are produced 
through gradual hydrogenation of gas-phase molecules ~\citep{tielen, hasegawa, herbst1,schutte}. 
Methanol is formed in such a manner, and its surface formation has received much attention over the years,  both from
theoreticians  and modelers \citep{tielens,herbst2,awad,rob}
and experimentalists \citep{hiraoka,watanabe1, watanabe2}. 

In order to simulate all of the chemistry that occurs in a cold region, it is necessary to include both the chemical kinetics in the gas and on the surfaces of dust grains.  The basic method for simulating chemistry is to use coupled rate equations, one for  each species in the simulation.   
Rate equations have been proven to be an accurate and efficient approach to simulate gas phase 
chemical kinetics \citep{herbst1,millar,wakelam}.  In these equations, the average concentration of a species is progagated forward in time through a non-linear differential equation that contains both source and sink terms corresponding to chemical reactions.  
Their success comes from the fact that average concentrations are normally well defined in the gas phase as functions of time.  In other words, 
 the so-called mean-field approximation is valid.  However, a straightforward and efficient generalization to simulate diffusive 
surface chemistry on interstellar grains  in a similar manner \citep{pickles,hasegawa} can be in error  \citep{tielen}.  
In particular, when the number of surface
species on an average grain is small, the rate equation method breaks down since its mean-field 
approach does not accurately describe the surface kinetics, which must be described by a method in which both the discrete number of species and fluctuations in this number are considered.
Several alternative methods to rate equations have been proposed to overcome this difficulty. 
The simplest of these is the modified-rate-equation approach,
which uses the rate equation method but instead of using the actual diffusion rate on the surface, scales this rate downward so that the rate of reaction does not on average exceed the rate of accretion of a reactant \citep{caselli}.  This semi-empirical approach is efficient computationally, but it is not 
reliable under all conditions \citep{herbst2,garrod}.

There are two more rigorous approaches, both of which are stochastic and macroscopic in nature.  In other words, they allow a determination of the probability that a certain number of molecules of a given species exists on a grain at any time without distinguishing where on the grain the adsorbate might lie.  In the Monte Carlo approach \citep{charnley1}, random numbers are used to determine what processes occur in a time interval based on their relative rates \citep{caselli2}.  It is difficult to develop a combined gas-grain model, in which the Monte Carlo method is used for the surface chemistry and the rate-equation approach for the gas-phase chemistry. The
 master equation approach \citep{biham,green,charnley}, on the other hand, is based on integrating a set of ordinary differential equations for the probabilities of surface populations,
and can be coupled with rate equations for gas phase chemistry.  Indeed, a coupling of the master equation approach to surface chemistry and a Monte Carlo approach to gas-phase chemistry has also been suggested \citep{charnley1}. Although the master equation approach
is believed to be the better candidate to solve the difficulties with the surface rate-equation 
approach, it has the drawback that there are many simultaneous equations to be solved, since the probabilities of surface populations for different species can be correlated. Nevertheless, with suitable approximations, two groups have successfully run gas-grain networks of cold interstellar cloud cores with
the master equation approach to model surface chemistry \citep{herbst2,biham2}. In these simulations, only a small subset of the surface reactions used in full-size models \citep{hasegawa} were included, with methanol the most complex species synthesized.

Recently, we adopted a microscopic stochastic method for surface chemistry, known as continuous-time random walk (CTRW) Monte Carlo  approach \citep{montroll}, which tracks
the specific trajectory and rate of diffusion of each adsorbate and the binding (desorption or evaporation) energy and barrier to diffusion at each local site \citep{chang,cuppen}. 
In this method, grain surfaces are modeled as square lattices while
hopping, evaporation and deposition are regarded as Poisson processes. 
We developed the CTRW approach for several reasons.
First, for a surface with a continuous distribution of diffusion barriers and evaporation energies, neither
the rate equation approach nor the macroscopic stochastic approaches can be easily implemented.  Even for a surface with discrete values of the energy parameters, known as a rough surface, the earlier approaches are not readily useable.  Moreover, it is facile to consider time-dependent barriers and desorption energies with the CTRW approach. Secondly, all previous methods neglected the "back diffusion"
of hopping species on a granular surface.  This neglect, however, turned out to be relatively unimportant for astronomical purposes \citep{krug,chang2}.  
Finally, the use of a microscopic Monte Carlo method allows us to follow the morphology of a mantle as it develops, so that only reactions in nearby monolayers can react with one another.  For instance, a hydrogen atom on the surface is not likely to react with an oxygen atom deeply buried
in the ice mantle. This constraint has been neglected in almost 
all previous gas-grain network calculations. 

Although the CTRW Monte Carlo approach allows us in principle to study both the detailed microscopic chemistry occurring in a grain mantle and the actual morphology of the mantle, use of the method presents some formidable computational  challenges.  Up to the present, we have used the method only to study the details of molecular hydrogen formation on surfaces relevant to diffuse clouds \citep{chang,cuppen,chang2}.   In a work in preparation, we are expanding our use of the method to study the formation of mantles of water ice and related species when hydrogen and oxygen atoms accrete onto a grain surface \citep{cuppen2}.  This extension still does not represent a true gas-grain model because the gas-phase chemistry is not followed.  Indeed,  there is no straightforward way to implement varying gas-phase abundances at the same time the surface chemistry progresses,  which makes it difficult to use the CTRW approach simultaneously with a gas phase kinetics simulation program.  

The difficulty with time-varying gas-phase abundances is that these lead to time-varying fluxes of accreting species onto the grains.  The CTRW method calculates deposition as a Poisson process with a time interval $\Delta t$ between events calculated from a random number $X$ between 0 and 1 determined via the relation $\Delta t=-\bar{t}\ln X$, where $\bar{t}$ is the 
average time interval between deposition. 
If abundances change, this equation would be invalid since $\bar{t}$ would no longer be  constant and the deposition would not be a strictly defined Poisson process.  This problem has been solved by \cite{jansen} for the simulation of a laboratory procedure known as temperature-programmed desorption (TPD), in which the temperature changes, resulting in a change of residence time for surface species. This technique, however,  cannot be directly applied to a gas-grain chemical model
because, unlike  the TPD simulation,  where the temperature is linearly dependent on time,  no prior knowledge of the abundances of gas-phase species  is available.      

In the present paper, we present a first attempt to combine a treatment of surface chemistry by the CTRW approach with a treatment of gas-phase chemistry by the standard rate equation approach. The iterative  method is based on rather weak coupling between the two chemistries, a condition that occurs primarily for cold cores, where thermal evaporation is negligible for most surface species.  The present paper uses a variant of the so-called  H, O and CO system as the surface reaction network \citep{charnley,herbst2}.  In this system, the gas-phase species atomic hydrogen, atomic oxygen, and carbon monoxide accrete onto grains and react to form species such as molecular oxygen, water, carbon dioxide, formaldehyde, methanol, and assorted radicals. The surface network, which is complex enough to mimic a full 
surface chemical reaction network, has  
been extensively studied by the modified rate
equation approach \citep{caselli2}, the macroscopic Monte Carlo method \citep{caselli2},
and the master equation approach \citep{herbst3} with fixed gas-phase abundances of the accreting species.  The master equation has also been used in a calculation, like the present, where gas-phase chemistry also occurs \citep{herbst2,biham2}.   By carefully studying the influence
of the desorption of species from the grain surfaces on the overall gas phase abundances, we are led to our weak coupling hypothesis that the gas-phase kinetics is almost independent 
of the surface chemical kinetics.  The weak coupling allows us to handle the chemistry in an iterative manner. 
        
The paper is organized as follows. Section 2 reviews the H, O and CO system, while Section 3 discusses the initial conditions and basic assumptions used in our treatment. The details of our method of simulation and 
surface model are discussed in Section 4. We present and discuss our results in Section 5, including a comparison with observations of molecular ices in cold regions.  Our conclusions are stated in Section 6.

\section{The H, O, and CO system}  
In the stochastic simulations to be discussed later,  we follow only the accretion of H, O and CO from the gas-phase and the surface reactions that result from this accretion. As will be discussed in detail, the accretion of all species is considered in a calculation prior to the Monte Carlo simulation.

Table \ref{ta1} shows the limited surface reaction network used in our stochastic simulations.  With two exceptions, the reactions can occur through the diffusive (Langmuir-Hinshelwood) mechanism until the reactants find one another in the same site.   The Eley-Rideal mechanism, in which a gas-phase species lands atop an adsorbed one and reacts with it, is also considered.  Although the formation of molecular hydrogen via recombination of hydrogen atoms is listed, as we discuss below, this process need not be considered if one starts with hydrogen in its molecular form. Most of the reactions in Table \ref{ta1} occur without activation energy $E_{\rm a}$ barriers;  for those that do possess activation energy, we use values taken from \cite{caselli2}. 
The H, C, and CO network in Table \ref{ta1} differs somewhat from that used by \cite{herbst2}.  These authors included two reactions to form carbon dioxide  -- the association reaction between CO and O and the reaction between O and the radical HCO -- which are  relatively  unimportant in our calculations  mainly because the large activation energy cannot be efficiently tunneled under by the heavy reactants. 
We also added a few reactions to make the system more realistic.
The most important reactions in this network concern the gradual hydrogenation of CO to produce
methanol, which has been studied in the laboratory  \citep{watanabe1,Fuchs:prep}.

In our simulations, which pertain to a source at a temperature of 10 K or 15 K, we assume that only H and O can hop,  and ignore the movement of all other species.  The rate of hopping depends exponentially on the energy barrier between adsorption sites; this barrier is significantly lower for these physisorbed atoms than for the other surface species followed.  At higher temperatures, the hopping of heavier species begins to become important  \citep{garrod}.  Likewise, thermal evaporation of almost all species is unimportant because it depends exponentially on the desorption energy, and this energy is much too large for species heavier than hydrogen at 10-15 K.  In fact, we need only follow the evaporation of hydrogen atoms.
Both the hopping barrier and the evaporation energy for H and O atoms are not constant
during our simulation.

Since we follow the morphology of the growing mantle, we find that among monolayers that lie next to one another, products of a reaction in one of the monolayers may lie atop  other reactive species in the next lower monolayer and be able to react with them.  The OH + H$_2$CO $\rightarrow $ HCO + H$_2$O reaction in Table \ref{ta1} is a good example.
The atom O cannot react with H$_2$CO, but it reacts with H to form OH.
If an H$_2$CO molecule  lies below the original oxygen atom,
the newly-formed OH product will directly react with H$_2$CO. What is interesting
is that OH and H$_2$CO are assumed to be stationary under these conditions,
but they can react by the diffusion of H atoms.

\begin{table}

\parbox{8cm}{\caption{Surface reactions in the H, O and CO system }\label{ta1}}\\
\begin{tabular}{lll}\hline \hline
Number & Reaction & $E_{a}$(K) \\ \hline
1 & H + H $\rightarrow$ H$_2$ & \\
2 & H + O $\rightarrow$ OH &  \\
3 & H + OH $\rightarrow$ H$_2$O & \\
4 & H + CO $\rightarrow$ HCO & 2500 \\
5 & H + HCO $\rightarrow$ H$_2$CO & \\
6 & H + H$_2$CO $\rightarrow$ H$_3$CO & 2500 \\
7 & H + H$_3$CO $\rightarrow$ CH$_3$OH & \\
8 & O + O $\rightarrow$ O$_2$ &  \\
9 & O + OH $\rightarrow$ O$_2$ + H &  \\
10 & CO + OH $\rightarrow$ CO$_2$ + H & 176 \\
11 & OH + H$_2$CO $\rightarrow$ HCO + H$_2$O & \\
12 & O + H$_3$CO $\rightarrow$ H$_2$CO + OH & \\ \hline
\end{tabular}
\end{table}

\section{Initial conditions and some basic assumptions} 
It is common in pseudo-time-dependent models of dense cloud chemistry to assume that the initial form of hydrogen is molecular, having been formed on grain surfaces during the diffuse cloud stage or somewhat later as the dense core was formed.   In dense cloud cores, H$_2$ is consumed to some extent
by the gas phase chemistry.  Indeed, previous model calculations at 10 K  show that cosmic ray ionization of H$_{2}$ leads to a residual atomic abundance of atomic hydrogen of $\approx$  1 cm$^{-3}$ while that for H$_2$ is maintained at 10$^4$ cm$^{-3}$.  But does this maintenance depend on the continuing surface formation of H$_{2}$ under dense cloud conditions?  Assuming a standard grain size of 0.1 $\mu$m and a standard gas-to-dust ratio, a sticking coefficient of unity for H atoms, and unit efficiency for the conversion of H to H$_{2}$ on grain surfaces, we obtain that at most 10\% of the initial H$_{2}$ abundance can be produced within 10$^{8}$ yr.  For a more reasonable time scale of 10$^{6}$ yr, the amount of H$_{2}$ produced is 100-fold less.
Despite this minimal production,
no depletion of H$_2$ has ever been reported even at higher temperatures when
the formation efficiency drops significantly, which indicates that gas-phase
chemistry during the dense cloud stage does not change the abundance of H$_2$ significantly.
So, by this argument, molecular hydrogen formation on dust grains in dense clouds can be neglected,
because it does not affect the gas.

In order to confirm the unimportance of H$_{2}$ formation and to understand the strength of the coupling between gas-phase chemistry and surface chemistry, we performed two calculations at 10 K using the rate equation approach.  In the first calculation, designated Model 1, we used the osu.2005 gas-grain code \citep{rob}, which includes over 4000 gas-phase reactions (see http://www.physics.ohio-state.edu/$\sim$eric/research.html)  and hundreds of surface reactions, both of which are treated by the rate-equation method.   Although all gas-phase species were allowed to accrete with unit sticking efficiency, we included only those surface reactions in Table \ref{ta1}.  In addition, atomic hydrogen was allowed to evaporate in competition with reaction.   In the second calculation, designated Model 2, we removed all the surface reactions and ran the network again with the following additional changes:  atomic hydrogen was not allowed to evaporate, and
the accretion of H$_2$ and He onto grains was halted, because we assume that these two species simply
land on grains and quickly evaporate again.

The initial fractional abundances for both models are given in Table \ref{ta2}. These are based on elemental abundances considered by \cite{garrod} based on unpublished work of Wakelam and Herbst.  As compared with more standard low-metal elemental abundances, there is in general less depletion of the heavier elements.   The constant density used is 
$n_{\rm H} $=$2 \times 10^4$ cm$^{-3}$ and the cosmic ray ionization rate $\zeta$ is set to $1.3 \times 10^{-17}$ s$^{-1}$.
We will use the same initial abundances and total density throughout the paper. Table \ref{table3} shows the comparison of fractional
abundances of selected species for these two models at a time of $10^6$ yr.
We can see that the difference is very limited; only the atomic hydrogen abundance is affected  significantly; its abundance is greater in Model 1 because evaporation is allowed.  But the extra atomic hydrogen does not make a significant difference for other species.  
This agreement shows that  the gas-phase
abundances remain mostly the same regardless of whether or not surface chemistry~\citep{herbst2} is considered as long as the temperature is sufficiently low that evaporation is not efficient for heavy species and non-thermal desorption is not included.   This latter constraint is an important one because non-thermal desorption of surface ices may indeed be non-negligible for a few gas-phase species such as methanol.  Nevertheless, in this first attempt to wed the Monte-Carlo microscopic  approach with gas-phase chemistry, we ignore the process and focus attention on the surface abundances.
Under these conditions, we can infer that the details of  surface kinetics are relatively
insignificant for  gas-phase abundances, whether or not they are handled by the rate-equation approach or by our more quantitative stochastic technique. The only important influence of grains
on gas-phase chemistry occurs by the accretion of species onto grains,  which eventually reduces
the abundance of gas-phase species.

\begin{table}
\parbox{8cm}{\caption{Initial Elemental Abundances}\label{ta2}}\\
\begin{tabular}{lc}\hline \hline
Species  & $n_i$/$n_{\rm H}$ \\
\hline
He       & 0.09\\
O        & $3.2 \times 10^{-4}$\\
H        & $5.0 \times 10^{-5}$\\
H$_2$    & 0.5\\
C        & $1.4 \times 10^{-4}$\\
S        & $1.5 \times 10^{-6}$\\
Si       & $1.95 \times 10^{-6}$\\
Fe       &  $7.4 \times 10^{-7}$\\
Na       &  $2.0 \times 10^{-8}$\\
Mg       &  $2.55 \times 10^{-6}$\\
P        &  $2.3 \times 10^{-8}$\\
Cl       &  $1.4 \times 10^{-8}$\\  \hline 
\end{tabular}
\end{table}

\begin{table}
\parbox{8cm}{\caption{Selected calculated fractional abundances at 10$^{6}$ yr}\label{table3}}\\
\begin{tabular}{lll}\hline \hline
Species & Model 1              & Model 2 \\
\hline
H       & $3.6\times 10^{-4}$  & $5.8\times 10^{-5}$ \\
H$_2$   & 0.50                 & 0.50\\
CO      & $8.9\times 10^{-6}$  & $9.2\times 10^{-6}$\\
O       & $1.1\times 10^{-5}$  & $1.1\times 10^{-5}$\\
C$_2$      & $3.7\times 10^{-10}$ & $4.4\times 10^{-10}$ \\
HCN     & $6.1\times 10^{-10}$ & $7.5\times 10^{-10}$\\
NH$_3$     & $4.2\times 10^{-9}$  & $4.3\times 10^{-9}$\\
H$_2$S     & $1.8\times 10^{-12}$ & $2.0\times 10^{-12}$\\
OH      & $1.6\times 10^{-9}$  & $1.8\times 10^{-9}$\\
C$_3$H     & $4.1\times 10^{-11}$ & $4.5\times 10^{-11}$\\ \hline
\end{tabular}

\end{table}

\section{Simulation Method and Surface Models}

\subsection{Calculation of Gas-Phase Chemistry}
\label{rhocalc}
Since the  two different approaches to surface chemistry give very similar gas-phase abundances
except for atomic hydrogen, we can start with a rate-equation approach to the gas-phase abundances coupled with accretion of all species onto grains with unit sticking efficiency and no subsequent evaporation of atomic hydrogen, as in Model 2.  
The assumption that evaporation of H does not occur at all derives from the work of 
\cite{chang} and \cite{cuppen}, who showed that at a
grain temperature of 10 K the recombination efficiency for molecular hydrogen is quite high for
realistic surfaces.  Nevertheless, it is not obviously true for all surfaces (see Table \ref{table3}), and
iteration with the actual Monte Carlo algorithm will be used to achieve better results.

The error in the assumption that hydrogen atoms do not return to the gas phase after accretion
can be calculated as follows.
Suppose that $F_{evap}(t)$ is the true time-dependent
rate of atomic H evaporated from a grain.  
The evolution of the gas-phase abundance of atomic hydrogen $n_g(H)$ can then be written as
\begin{eqnarray}
\frac{dn_g(H)}{dt}&=& \sum {\rm{formation}}^{gas} -\sum {\rm{destruction}}^{gas} \nonumber \\
                               &  &  -k_{acc}n_g(H)+F_{evap},
\end{eqnarray}
where $k_{acc}$ is the accretion rate coefficient and the first two terms on the right-hand-side of the equation refer to gas-phase formation and destruction processes. Formal integration of this equation yields
\begin{eqnarray}
n_g(H)(t)&= & \int \sum {\rm formation}^{gas} (t^{'}) \rm{d} t^{'} \nonumber\\
                 &   & - \int \sum {\rm destruction}^{gas}(t^{'})\rm{d}t^{'} \nonumber \\
          &   & -(1-\rho)\int k_{acc}n_g(H)(t^{'}) \rm{d}t^{'},
\end{eqnarray}
where
\begin{equation}
\rho=\frac{\int F_{evap}(t^{'}) \rm{d}t^{'} }{\int k_{acc}n_g(H)(t^{'}) \rm{d}t^{'}},
\end{equation}
represents the ratio between the integrated evaporation rate and accretion rate of atomic hydrogen.  This ratio can be easily determined during a Monte Carlo simulation, as discussed later, by counting the number of evaporating and accreting atoms over the period of time.
If $\rho$ is much less than one, the error is small and the
approximation of assuming $F_{evap}(t)$ to be zero is justified. If, however, $\rho$ is large,
the abundance of gas-phase atomic hydrogen 
can have a large error.  Although this error will not affect other  gas-phase abundances substantially, the incorrect hydrogen abundance will not be suitable for the subsequent surface chemistry simulation, which depends on the accretion flux of H atoms.
In this case, an iteration has
to be performed to obtain the correct gas-phase H abundance as a function of time.
As a first-order approximation, we assume  the ratio of evaporating
H to incoming H to be constant throughout the simulation and equal to $\rho$ determined over the entire time scale of the simulation.
This approach corresponds to taking the average evaporating
effect into account.   It is undertaken because the Gear routine used for the gas-grain calculation cannot solve ordinary differential equations with variables explicit in time. In other words, a variable $1 - \rho(t)$ factor cannot be inserted into the current gas-grain code.  After $\rho$ is obtained from the Monte Carlo simulation,
we reduce the deposition rate coefficient of atomic hydrogen, $k_{acc}$, to $k_{acc}(1-\rho)$ and run
the  simulation again. If the new value of $\rho$ is close to the previous one,
the system is converged, otherwise the cycle is repeated with the new $\rho$ until convergence is achieved.  

If $\rho$ has no
time dependence, the system converges to the true solution. However, $\rho$ is typically time dependent
and this time dependence cannot be included in the gas-grain network calculation. In order to gauge the
resulting error, we define a time-dependent $\rho_2(t)$, which is the ratio of the number of evaporating H atoms to
incoming H atoms counted during a ``short" time period. We choose this ``short" time to be $2 \times 10^3$ yr, which is
1/100 of the total simulation time of $2 \times 10^5$ yr.
The time is short enough to determine the time dependence of $\rho_2(t)$ and long enough
to suppress most of the statistical noise.  A shorter time scale ($ 1 \times 10^{3}$ yr) shows some more noise while a longer time scale ($5 \times 10^{3}$ yr) shows less.
With the rate equation approach, this method is equivalent to the integration
\begin{equation}
\rho_2(t)=\frac{\int_{t-2000~yr}^{t} F_{evap}(t^{'})\rm{d}t^{'}}{\int_{t-2000~yr}^{t} k_{acc}n_g(H)(t^{'})\rm{d}t^{'}}.
\label{rho2}
\end{equation}
If $\rho_2(t)$ is weakly dependent on time, $t$, then $\rho$ has little time dependence,
thus the resulting error is small. Otherwise, the error is large even if convergence has been achieved.
With the current Gear algorithm used in the gas-grain model, we cannot use the parameter $\rho_{2}(t)$ in the actual simulation.  A new gas-grain algorithm would have to be designed to follow the gas-phase abundance of H with the time-dependent $\rho_{2}$.

\subsection{Deposition of Gas-Phase Species}   
\label{dep}  
Assuming the sticking coefficients for each species to be 1, the
deposition (accretion) flux of species $i$ onto each absorption (lattice) site is equal to the product $f_{i}N$, where $f_i$ is the incoming flux of species $i$ in units of ML s$^{-1}$, while $N$,  the number of sites on the grain, is given by
\begin{equation}
N=4\pi r^2 s,
\end{equation}
where $r$ is the radius of the grain and $s$ is the surface site density.  In our simulations, we start with bare olivine as the granular surface, and use a site density
$s$ of  $1.5\times 10^{15}$ cm$^{-2}$, which is also used in our standard gas-grain network.
The flux $f_i$ can be calculated by
\begin{equation}
f_{i}=\frac{n_g(i)}{s}\sqrt{\frac{8 k_B T_{g}}{\pi m_i}},
\end{equation}
where $T_g$ is the gas phase temperature and $m_i$ is the mass of species $i$.     A factor of four, applicable to the case of spherical grains, is often used in the denominator.  However, this factor is not used in the accretion term in our standard gas-grain network, and for consistency we do not use it here.

If $f_i$ is independent of time, the deposition is  an homogeneous Poisson process, and the deposition interval $\Delta t$ can then be
calculated using 
\begin{equation}
\Delta t=-\frac{\ln(X)}{f_i N},
\end{equation}
where $X$ is a random number between 0 and 1.
In our simulations the deposition rate is time-dependent 
due to the time dependence of the gas-phase abundances of the impinging species. These gas-phase abundances are mean-field values, but the flux is still stochastic in arrival time.  In reality, there is already a distribution in the flux due to a distribution in velocity.  The next 
deposition time, $t_{next}$, satisfies the equation,
\begin{equation}
X=\exp\left(-\int_{t_{prev}}^{t_{next}} f_i(t^{'}) N \rm{d}t^{'}\right),
\label{X}
\end{equation}
with $t_{prev}$ the previous deposition time, and the process is referred to as an inhomogeneous Poisson one. 
It can easily be shown that Eq.~(\ref{X}) is also valid when $f_i$ is independent of time. 
With an analytical form
of $f_i(t)$ known, $t_{next}$ can be determined numerically. However, only a numerical solution is obtained from the
gas-grain network calculation, thus interpolation is needed to approximate $f_i$. The current gas-grain
network outputs gas phase abundance of species $i$, at different time $t_j$. We employ linear interpolation
to get the abundance of species $i$ at time $t$, which is between $t_j$ and $t_{j+1}$. 
The flux $f_i(t)$  can then be calculated as
\begin{equation}
f_i(t)=f_i(t_j)+\frac{f_i(t_{j+1})-f_i(t_j)}{\Delta t_j}(t-t_j),
\label{inte}
\end{equation}
where $\Delta t_j=t_{j+1}-t_{j}.$
We can formally write $f_i(t)$ for any $t$ as,
\begin{eqnarray}
f_i(t)&=&\sum_{j=0} (f_i(t_j)+b_{ij}(t-t_j))(S(t-t_j) \nonumber \\
        &   & -S(t-t_{j+1})),
\label{fi}
\end{eqnarray}
where $S(t)$ is the step function and $b_{ij}= \frac{f_i(t_{j+1})-f_i(t_j)}{\Delta t_j}.$  
Combining Eqs.~(\ref{X}) and (\ref{fi}) gives
\begin{eqnarray}
X&=&\exp(-\int_{t_{prev}}^{t_{next}} \sum_{j=0} (f_i(t_j)+b_{ij}(t^{'}-t_j)) \times  \nonumber \\
& &  (S(t^{'}-t_j)-S(t^{'}-t_{j+1})) N\rm{d}t^{'}).
\end{eqnarray}
Let us call 
\begin{eqnarray}
I(t_{prev},t_{next}) &\equiv& \int_{t_{prev}}^{t_{next}}
\sum_{j=0} (f_i(t_j)+b_{ij}(t^{'}-t_j)) \times \nonumber \\
& & (S(t^{'}-t_j)-S(t^{'}-t_{j+1})) \rm{d}t^{'}).
\label{I}
\end{eqnarray}
Then
\begin{equation}
r = I(t_{prev},t_{next}),
\end{equation}
where $r=\frac{-\ln X}{N}$.

Eq.~(\ref{I}) can be solved numerically in the following manner. Suppose $t_k\leq t_{prev} < t_{k+1}$.
First, we calculate $I(t_{prev},t_{k+1})$. If this is 
larger than $r$, we know that $t_{next}$ must be less than $t_{k+1}$.  We can determine $t_{next}$ by expressing $f_{i}(t)$ in terms of $t_{prev}$ rather than in terms of $t_{k}$. Integration then yields 
\begin{equation}
t_{next}=t_{prev}+\frac{\sqrt{f_i^2(t_{prev})+2rb_{ik}}-f_i(t_{prev})}{b_{ik}},
\label{solu}
\end{equation}
where $f_i(t_{prev})=f_i(t_k)+b_{ik}(t_{prev}-t_k)$. 

 If $I(t_{prev},t_{k+1}) < r$, then deposition
must happen later than $t_{k+1}$.  We define a new quantity, $r^{'}$,
\begin{equation}
r^{'}=r-I(t_{prev},t_{k+1})
\end{equation}
or
\begin{equation}
r^{'}=I(t_{k+1},t_{next} ).
\end{equation}
If integration from $t_{k+1}$ to $t_{k+2}$ leads to a larger value than $r^{'}$, $t_{next} < t_{k+2}$ and an expression for $t_{next}$ similar to eq.~(\ref{solu}) can be obtained.  If integration to $t_{k+2}$ leads to a smaller value than $r^{'}$,  $t_{next}$ is larger than $t_{k+2}$, and the sequence must be continued until $t_{next}$ is bounded by two successive times $t_{m}$ and $t_{m+1}$.  

  To check for convergence, the original results of the gas-grain network are rerun twice with smaller time intervals, first 1/4 of the original
$\Delta t_j$ and then 1/4 of the second one.  Convergence is achieved in all cases.

\subsection{Surface Models}
A detailed description of flat and rough surface models has been given in 
\cite{cuppen}. Here we start with either a flat surface or their
surface D, which is the roughest one and is generated by another Monte Carlo program. The diffusion barrier, $E_b$,
and the desorption energy, $E_D$, for the two diffusing species -- atomic hydrogen and atomic oxygen -- are in our models dependent on the topological roughness.
Starting from either a flat or rough bare olivine surface, a mantle gradually develops.
The formation of surface species, especially water ice, on the surface has two consequences.
First, the chemical nature of the surface changes, resulting in
different energetics and therefore different values for $E_b$ and $E_D$. Secondly, the
surface topology changes, as more and more species are formed  and roughness develops even from the initially flat surface due to fluctuations.
Thus, $E_b$ and $E_D$ are inherently stochastic in the simulation in that they change for the site at the top of a given column as the stochastic calculation proceeds.
For simplicity, we assume that only the olivine substrate and the H$_2$O molecules (the main mantle constituent)
influence $E_b$ and $E_D$, either by lateral interactions or by a vertical interaction
with the substrate. 

The parameters $E_b$ and $E_D$ are given by
\begin{equation}
E_b=E_b^0+iE_L^1+jE_L^2,
\end{equation}
\begin{equation}
E_D=E_D^0+iE_L^1+jE_L^2,
\end{equation}
respectively. Here $E_b^0$ and $E_D^0$ are the diffusion barrier and desorption energy on a
 surface without lateral bonds, while  $E_L^1$ and
$E_L^2$ are the lateral bonds imposed by horizontal olivine and H$_2$O neighbors respectively.
In this simulation, $E_L$ has a value of
$0.1E_D$~\citep{cuppen}. The indices $i$ and $j$ are the number of olivine and H$_2$O neighbors respectively. Since species
other than H and O have large diffusion barriers even without lateral bonds~\citep{ruffle}, 
we ignore their movement in the temperature range studied here (10-15 K). The evaporation of O is ignored.

Table \ref{table4} summarizes
the parameters used for H and O on the ice and olivine surfaces.
The diffusion barrier and desorption energy of H on olivine are from \citet{katz}.
The desorption energy for H on water ice is not well constrained.
\citet{hollen} found a value of 450 K, \citet{buch} found
a value of 500 K, while \citet{alha} found a value of 400 K.
We take the average of these measurements.  
Recently, \citet{perets} deduced a much higher value  of 720 K.
The oxygen values are from \citet{tielen}.  
For comparison, we also simulate a second
 system by ignoring the lateral bonds and only considering the change of the chemical nature of the surface,
i.e. the vertical bond.  This system starts from a flat surface.  Although it gains some roughness from fluctuations as the ice layers grow, the  system is kinetically equivalent to a flat system since there is no penalty for going uphill and no gain in falling into a valley.

\begin{table}
\caption{ Energies (K) used for different surfaces}
\label{table4}
\begin{tabular}{lll}\hline \hline
Paarameter         & Olivine     & Ice \\ \hline
$E_{b,H}^0$    & 287     & 346 \\ 
$E_{b,O}^0$    & 616     & 616 \\ 
$E_{D,H}^0$    & 373     & 450 \\ 
$E_{D,O}^0$    & 800     & 800 \\ 
$E_{L,H}$      & 37.3    & 45 \\ 
$E_{L,O}$      & 80      & 80 \\ \hline
\end{tabular}
\label{table4}
\end{table}

\subsection{Surface reactions with activation energy barriers}

In gas-grain codes based on a rate equation or master equation approach to the surface chemistry,
the common method to calculate diffusive rate coefficients for reactions with an activation energy
barrier is to multiply the diffusion rate by a tunneling factor:
\begin{equation}
k_{ij}=\exp\left(-\frac{2a}{h}\sqrt{2 \mu E_a(i,j)}\right)
\end{equation}
where $a$ is the width of the rectangular barrier, typically chosen to be 1 \AA,
$\mu$ is the reduced mass, and $E_a$ is the reaction activation energy \citep{hasegawa}.  Note that the word ``diffusion'' is normally used to refer to motion over a whole grain, whereas the word ``hopping'' refers to motion from one potential minimum to the next.
A physical interpretation of the rate coefficient is that the two species have a probability equal to the tunneling
expression to react with each other if they diffuse into the same potential minimum. Thus, the rate coefficients
are linearly dependent on the diffusion rate.

Unlike the typical situation in the gas phase, two species that meet each other on a surface can stay in each other's vicinity for quite a while
depending on the hopping rate to nearby lattice sites.  During this time they can have many chances to react. 
It is therefore more physical to model reactions with barriers as processes competitive with hopping out of the potential minimum.  Such an approach has been done by \cite{awad} using rate equations.  Here we incorporate this approach by considering the competition  among Poisson processes. 

In particular, the competition among reaction, hopping or evaporation can be implemented as follows. 
The reaction rate with tunneling can be written as
\begin{equation}
b_{r}=\nu_1 \exp\left(-\frac{2a}{\hbar}\sqrt{2\mu E_a(i,j)}\right) ,
\end{equation}
where $\nu_1$ is the attempt frequency for reaction.  The competing rates for hopping and evaporation of the two species are given by
\begin{equation}
b_1=\nu_2 \exp\left(-E_{b}/T\right) ,
\label{b1}
\end{equation}
and 
\begin{equation}
b_2=\nu_3 \exp\left(-E_{D}/T\right) ,
\label{b2}
\end{equation}
respectively, where $T$ is the surface temperature, $\nu_2$ is the attempt frequency for hopping, and 
$\nu_3$ is the attempt frequency for evaporation, and the diffusion barrier and evaporation energy pertain to either of the two species.  The stochastic algorithm for diffusion treats all directions with equal probability.  Note that hopping is treated classically rather than by quantum mechanical tunneling because, unlike reaction, hopping is thought to occur over a broad and shallow potential.  Quantum chemical calculations are being undertaken to verify this point (Woon, private communication).
Assuming that only one species is moving,  the probability for reaction to occur instead of diffusion or evaporation is
\begin{equation}
p=\frac{b_r}{b_r+b_1+b_2}.
\label{p}
\end{equation}
We assume $\nu_1=\nu_2=\nu_3 =10^{12}$ s$^{-1}$,
but since there three different types of processes the attempt frequencies can be different.
The competition is implemented in the Monte Carlo algorithm as follows: a random number between 0 and 1 is generated.
If this number is smaller than $p$ the reaction occurs.  Otherwise evaporating or hopping takes place. A competition between these two processes must also be evaluated.

\subsection{Monte Carlo Simulation}
The Monte Carlo algorithm starts with determining the first deposition time for H, O, 
and CO, as explained in Section \ref{dep}.  Then the first deposition is executed by placing the particle 
on a randomly picked site of the lattice, which is chosen to have a lattice of size  $100\times 100$.  This size corresponds to a rather small grain of 0.015 $\mu$m,  and a mean distance between atoms in adjacent lattice sites of $\approx 5 \AA$.  Typical activation energy barriers are shorter than this in width;  the widths of the diffusion barriers are currently under investigation (Woon, private communication).  We have already shown that the lattice size used is large enough that there is little size dependence to the results for both flat and rough olivine 
surfaces \citep{chang,cuppen}.     Nevertheless, we redid our simulations for lattice sizes of $50 \times 50$ and $200 \times 200$ lattices at 15 K to confirm this point.  For much smaller grains, the so-called accretion limit is reached, where the average mantle abundances of reactive species is less than unity, leading to a drop in reaction efficiency.   For everything but the surface stochastic simulations, the dust-to-gas number density is taken to be  $1.32\times 10^{-12}$ and the grain size the nominal one of radius  0.1$\mu$m.

As the clock moves forward, other species are deposited on the lattice, and species already on the lattice can hop from site to site or evaporate.   By hopping on a rough surface, we mean motion on the top level of any occupied site, whether is is occupied by the original silicate or by any other species.  Thus hopping can include climbing or falling in a vertical sense.  For very rough surfaces, the climbing or falling length can be up to 20-30 molecules. 
Evaporation and hopping are assumed to be Poisson distributed, with time intervals between events
$\Delta t=-\frac{\ln(X)}{b}$, where $X$ is a random number between 0 and 1, and $b$ is the occurring
rate of that process as given by Eqs.~(\ref{b1}) and (\ref{b2}).  For H atoms, which are allowed to both hop and evaporate, these two processes are treated first
as one combined process with a rate $b_1+b_2$ and then another random number $X^{'}$ is called to decide which one will occur first. If $X^{'}$ is smaller than
$\frac{b_1}{b_1+b_2}$, the hydrogen atom hops and it evaporates in the other case. The total number of H atoms that evaporate is
counted to calculate $\rho$.   
Oxygen atoms are only allowed to hop. The time interval between hopping is solely determined by the hopping rate.  

If a hydrogen or oxygen atom hops to a site where there is already one species that can react with it without a barrier, the reaction happens immediately.  If either atom hops into the same site as a species with which it can react with an activation energy barrier,  the competition among reaction and other processes occurs as discussed in the previous section. 
Reaction products other than H$_2$ remain on the surface
and can cover  species on the same sites.  
Some reaction products can react further with species they cover. For instance, if OH is produced by a reaction in which H hops into a site occupied by an O atom, 
and there is an H$_2$CO molecule lying under the O atom, then another reaction is allowed to take place, producing HCO and H$_2$O (see Table~\ref{ta1}).

If a gas-phase species lands atop a bound surface species, reaction can occur according to the Eley-Rideal mechanism.    No distinction is made between reaction by hopping (Langmuir-Hinshelwood mechanism) and by landing.   If reaction does not occur upon landing, the incoming particle covers the one already present.  No reaction is allowed between species covered by at least one other species.  For instance, assume that an oxygen atom is buried by a CO molecule.  After being buried, the oxygen atom cannot react with H or O even if they occupy the site above the CO.  

 The Monte Carlo simulation is performed by executing the following sequence of steps:\\
1)  The gas-grain code is run for Model 2 through 10$^{6}$ yr,\\
2)  The gas-phase results for H, O, CO are used to run the Monte Carlo simulation of deposition, surface chemistry, and evaporation of atomic hydrogen for $2\times 10^{5}$ yr,\\
3)  An average value of $\rho$  over the full time of the stochastic calculation is obtained,\\
4)  If $\rho$ is close to zero or is converged, the calculation is stopped, otherwise:\\
5)  The gas-grain code is rerun with a deposition rate reduced by a factor of 1 - $\rho$, and the simulation is also run with a reduced deposition rate for H,\\  
6)  The procedure then returns to step 3.\\

Note that the gas-grain code is used to obtain gas-phase results; surface results are subsequently obtained by stochastic simulation.  The Monte Carlo code is run for a shorter time than the gas-grain code for two main reasons: (i) it is very time-consuming computationally, and (ii) the parameter $\rho_{2}(t)$ becomes significantly time-dependent after $2 \times 10^{5}$ yr.  Finally, the iterative procedure used is forced on us by the current structure of the gas-grain code, which does not allow parameters with explicit time dependence.

\section{Results and Discussion}

\begin{figure}
\centering
\rotatebox{-90}{\includegraphics[width=20
em]{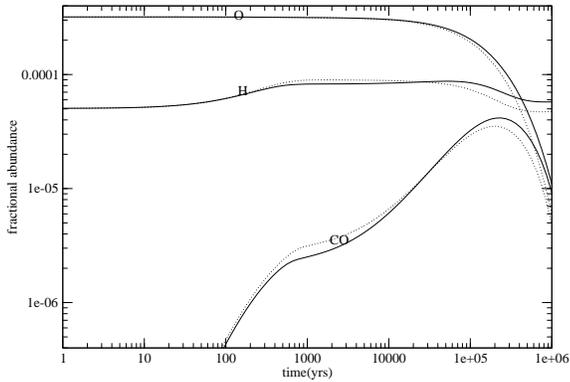}}
\caption{Fractional abundances of gas-phase species that accrete onto grains for Model 2.
Solid lines are for 10 K while dotted lines are for 15 K.}
\label{fig1eh}
\end{figure}
The simulation starts in earnest with runs at 10 K and 15 K in which the fluxes of hydrogen atoms, oxygen atoms, and CO striking the grain as a function of time are taken from Model 2. 
 Figure \ref{fig1eh} shows the fractional abundances of gaseous H, O and CO as functions of time at 10 K and 15 K obtained from the Model 2 run.  As can be seen, the atomic hydrogen abundances do not change much over the first 10$^5$ yr, while the decline in the atomic oxygen abundance turns on after 10$^{4}$ yr.   A significant amount of gas-phase CO is produced after 10$^{3}$ yr.
 These abundances are those used to calculate fluxes for the Monte Carlo simulation of the surface chemistry.  In this run, both surface chemistry and evaporation of atomic hydrogen occur.  As discussed in Section \ref{rhocalc}, the evaporation of atomic hydrogen is followed stochastically and averaged over time to determine the parameter $\rho$.  If this parameter is significantly different from zero, then further runs are necessary with a reduced hydrogen flux to achieve convergence. The closer the value of $\rho$ to unity, the slower the convergence.

We first simulate the surface chemistry  starting with a flat olivine surface and using a surface-adsorbate system with no lateral bonds as the ice develops.  We then redo
the simulation starting from a rough olivine surface and containing lateral bonds.  The latter surface, which is rough at all stages,  will just be referred to as ``rough''.

\subsection{Surface without lateral bonds}
\begin{figure}
\resizebox{8.0cm}{!}
{\includegraphics{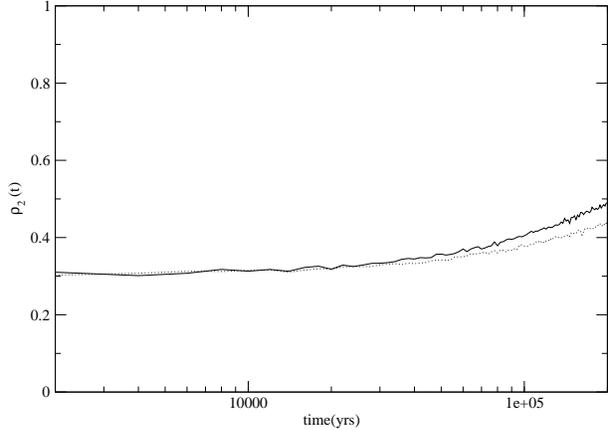}}
\caption{$\rho_2(t)$  vs time t for a  surface-adsorbate system without lateral bonds at 15 K.
The solid line is for $\rho_2$ after iteration, while the dotted line is for $\rho_2$ calculated in the first Monte Carlo simulation.
}
\label{fig2eh}
\end{figure}

At 10 K, $\rho$ is initially calculated to be  $3 \times 10^{-4}$, which is significantly less than 1. Thus, the gas-phase fractional abundances shown in Fig.~\ref{fig1eh} are also reasonably good solutions of the gas-grain network with evaporation.  
However, at 15 K, $\rho$ is 0.37, which is comparable to 1, because evaporation is more important. Following the prescription in Section \ref{rhocalc}, we reduce the hydrogen-atom accretion rate to 0.63 of the previous value, and perform the gas-grain simulation a second time with the Monte Carlo approach to the surface chemistry.  In this simulation, $\rho$ now is calculated to be 0.40, which indicates that reasonable average convergence has been reached.  Moreover,  we can look at the time-dependent $\rho_{2}(t)$ to check that it is reasonably independent of time, another test of convergence. 
Figure ~\ref{fig2eh} shows $\rho_2(t)$ as a function of time for 15 K.  
From this figure, we see that the error of our approach is small
since $\rho_2(t)$ converges and does not change much with time, although there is an increase after 10$^{5}$ yr.  
The increase of $\rho_2(t)$ derives from  the decrease of O in the gas phase, which decreases its flux onto grain surfaces, 
so that more H atoms can evaporate
before combining with surface O.

\begin{figure}
\centering
\resizebox{8.0cm}{!}
{\includegraphics{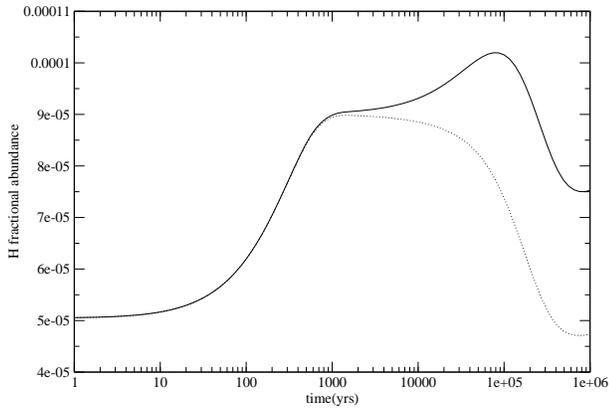}}
\caption{Fractional abundance of gas-phase atomic hydrogen at 15 K. The solid line refers to the first simulation with the Monte Carlo method while the dotted line refers to a subsequent iteration.
}
\label{fig3eh}
\end{figure}

The gaseous fractional abundance of atomic hydrogen after the initial Monte Carlo simulation and its iteration at 15 K is shown in Fig.~ \ref{fig3eh}.
As expected, the H fractional abundance
increases after the initial surface simulation since evaporation is allowed. For example, at 10$^{5}$ yr, the H fractional abundance exceeds 10$^{-4}$ whereas it is closer to $7 \times 10^{-5}$ for Model 2.  After the iteration, however, the atomic H abundance has become closer to its original abundance at all times.  Even with the increased gas-phase H abundance after the initial simulation, the abundance for each surface species changes less than 15\% at 15 K.
This indicates that the change in the
gas-phase atomic hydrogen abundance due to the change in H evaporation from the
grain surface has little effect on the surface abundances.
The other gas-phase abundances do not change much either because
atomic hydrogen only plays a minor role in the gas-phase chemistry.

\begin{figure}
\resizebox{8.0cm}{!}
{\includegraphics{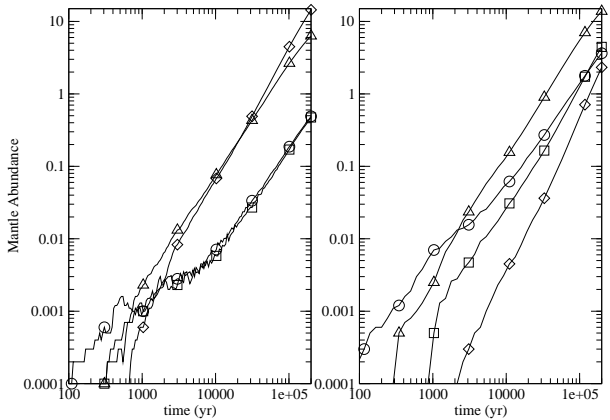}}
\caption{Mantle abundances for selected species in monolayers
as a function of time on a surface-adsorbate system with no lateral bonds. Circles are for CO, triangles pointed upward are
for CO$_2$, squares are for H$_2$CO, while diamonds are for CH$_3$OH. The left panel refers to results at 10 K while
the right panel refers to results at 15 K.
}
\label{fig4eh}
\end{figure}

Figure \ref{fig4eh} shows the surface abundances of CO, H$_2$CO, CH$_3$OH and CO$_2$
on grain surfaces as functions of time at 10 K and 15 K.
After some initial fluctuations, especially at 10 K,
the population of each carbon-bearing species (in monolayers per grain) increases monotonically with time. At 10 K, methanol exceeds CO after 500 yr
 and eventually becomes the most abundant carbon-bearing surface species, while  CO$_2$ is also abundantly produced.
At 15 K, on the other hand, methanol is always the least abundant of the  four carbon-bearing species shown. 

Table \ref{table5} shows the populations of all the mantle species  at a time of $2 \times 10^5$ yr.  It is easily seen that the mantles are dominated by water ice, with molecular oxygen second in abundance.  
The total number of species at 15 K is slightly
larger because of the larger accretion rate at higher temperatures and the still minimal evaporation rate.
As already seen in Fig.~ \ref{fig4eh}, the abundance of methanol is highly dependent on temperature; its abundance drops by a factor of seven as the temperature increases from 10 K to 15 K.  The higher temperature makes the formation of methanol less efficient because
diffusion rates increase quickly at higher temperatures while the
tunneling reaction rates for the critical reactions with activation energy -- H + CO and H + H$_{2}$CO  -- are independent
of temperature.  Thus, competition favors hopping of H over reaction as the temperature increases.   This decrease in the methanol abundance with increasing temperature is not exactly what is measured in the laboratory, although the conditions there are somewhat different from those studied here 
 \citep{Fuchs:prep}.  Our analysis for the calculated temperature dependence of methanol also leads to  more HCO and H$_3$CO at 10 K than at 15 K. In addition, the
increasing temperature makes hydrogen atoms hop
faster, thus the activation energy-less reactions  with HCO and H$_3$CO become more efficient.
For HCO and H$_3$CO, there is another mechanism for destruction:
O atoms become mobile at 15 K,
and can react with both radicals.
Unlike other simulations, much CO$_2$ is produced in our calculation, mainly because
 products from one reaction can
 react further with
species under them; thus, species that are not mobile can react with each other. In this case, the reaction of importance is that between CO and OH,  the barrier of which is low enough that the probability of reaction is near unity.
For the case of O$_{2}$, its formation by surface recombination of oxygen atoms is helped by an increase in temperature since the surface atoms move around significantly more quickly at 15 K.  At 10 K, there is a significant amount of atomic oxygen in the grain mantles,
most of which is buried and not able to react further.

\begin{table}
\parbox{12.0cm}
{\caption{Mantle populations (ML)  $2\times 10^5$ yr (no lateral bonds). }\label{table5}}\\
\begin{tabular}{lll}\hline \hline
Species   & 10 K    & 15 K\\
\hline
H         & 0      & 0  \\
O         & 0.25   & 0.0017 \\
O$_2$     & 35     & 50 \\
OH        & 0.049  & 0.0012 \\
H$_2$O    & 139    & 138 \\
CO        & 0.49   & 3.6 \\
HCO       & 0.25   & 0.0075   \\
H$_2$CO   & 0.47   & 4.5 \\
H$_3$CO   & 0.19   & 0.001 \\
CH$_3$OH  & 15     & 2.3 \\
CO$_2$    & 6.3    & 14 \\
Total     & 197    & 212 \\ \hline
\end{tabular}

\end{table}

\begin{figure}
\resizebox{8.0cm}{!}
{\includegraphics{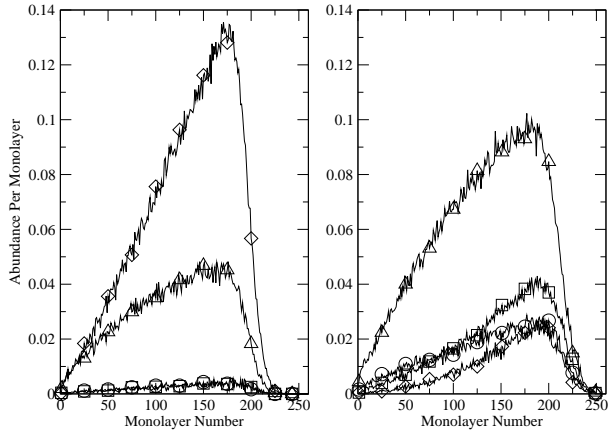}}
\caption{Abundance per monolayer in fractions of a monolayer for selected species 
at each layer on a surface without lateral bonds at $2 \times 10^{5}$ yr.  Monolayers are numbered from zero at the initial surface.   The symbols representing the species are the
same as in Fig.~\ref{fig4eh}. The left and right panels represent results at 10 K and 15 K, respectively.
}
\label{fig5eh}
\end{figure}

Figure \ref{fig5eh} shows the abundances of the major carbon-bearing species in individual monolayers, starting from the surface and moving upward. 
These differential abundances of each carbon-containing species typically increase initially as a function of height from the original surface  but peak before the
top layers are reached at both 10 K and 15 K. The bottom layers hardly have any molecules with carbon,
because gas-phase CO takes some time to produce. 
The top layers contain fewer species overall because these layers are not yet fully occupied.

\subsection{Rough surface}
The calculated $\rho$ at 10 K is about $1\times10^{-4}$, which is much smaller than unity, so that no iteration
is required. At 15 K, $\rho$ is 0.14, which is still smaller than 1. But we perform an iteration anyway. 
After one iteration, $\rho$ is still 0.14, so the calculation is converged.  

\begin{figure}
\resizebox{8.0cm}{!}
{\includegraphics{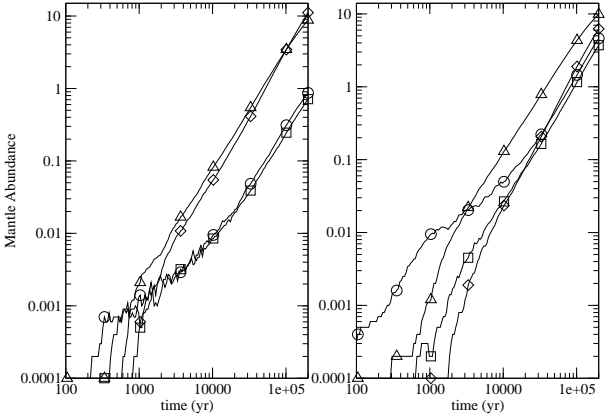}}
\caption{Mantle abundances for selected species in monolayers
as a function of time on a rough surface. Circles are for CO, triangles pointed upward are
for CO$_2$, squares are for H$_2$CO, while diamonds are for CH$_3$OH. The left panel refers to results at 10 K while the right panel refers to results at 15 K.
}
\label{fig6eh}
\end{figure}

The time-dependent surface abundances of CO, H$_2$CO, CH$_3$OH and CO$_2$ are shown in
Fig.~\ref{fig6eh}. At 10 K, the figure is similar to Fig.~\ref{fig4eh}, which represents the analogous results on a surface-adsorbate system without lateral bonds, while at
15 K, on the other hand, more methanol is produced than in the system without lateral bonds, because
there are stronger binding sites due to lateral bond interactions and these stronger binding sites
help to add hydrogen atoms to molecules when there is a reaction barrier,
since the competitive diffusion rate is lowered.

The surface abundances of all studied species at $2 \times 10^5$ yr are shown in Table \ref{table6}. 
Compared with Table \ref{table5}, there is not much difference at 10 K
except that more hydrogen atoms are trapped on the surface.  Indeed, for any reasonably-sized grain, the mantle population of 0.019 corresponds to far greater than 1 atom per grain.   At 15 K, however, as in Fig.~\ref{fig6eh}, one can see that more methanol is produced than without lateral bonds.

\begin{table}
\parbox{12.0cm}{\caption{Mantle populations (ML) $2\times 10^5$ yr  (rough surface)}\label{table6}}\\
\begin{tabular}{lll}\hline \hline
Species   & 10 K    & 15 K \\
\hline
H         & 0.019  & 0 \\
O         & 0.48   & 0.031 \\
O$_2$     & 46     & 41 \\
OH        & 0.073  & 0.0033  \\
H$_2$O    & 116    & 160 \\
CO        & 0.87   & 4.6 \\
HCO       & 0.38   & 0.025 \\
H$_2$CO   & 0.71   & 3.7 \\
H$_3$CO   & 0.26   & 0.0054 \\
CH$_3$OH  & 11     & 6.2 \\
CO$_2$    & 8.8    & 9.9 \\
Total     & 184    & 225\\ \hline
\end{tabular}

\end{table}

\begin{figure}
\resizebox{8.0cm}{!}
{\includegraphics{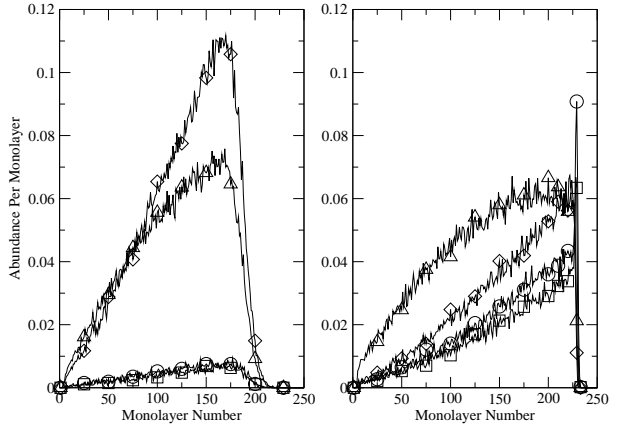}}
\caption{Abundance per monolayer in fractions of a monolayer for selected species 
at each layer on a rough grain surface at $2 \times 10^{5}$ yr.  Monolayers are numbered from zero at the initial surface.   The symbols representing the species are the
same as in Fig. ~5. The left and right panels represent results at 10 K and 15 K, respectively.
}
\label{fig7eh}
\end{figure}

The abundance per monolayer of the carbon-bearing species at different
layers is shown in Fig.~\ref{fig7eh} for a time of $2 \times 10^{5}$ yr.
For 10 K, this gives a very similar result to the system without lateral bonds in Fig.~\ref{fig5eh}. At 15 K,
more methanol is formed at each layer because of the slower hopping rates caused by the lateral bond interaction.

At the rough surface at 15 K,  the populations of species as a function of monolayer  have a distinct feature, in that they drop to zero
abruptly, unlike the other cases.
This effect happens because the oxygen atoms are able to  hop efficiently out of sites with small
lateral bond interaction, located above surrounding sites, to sites  with larger
lateral bond interaction, located in valleys below surrounding sites, where they are trapped.   So, sites that are lower in vertical height above the original surface tend to have a larger
probability to be occupied by oxygen atoms. These  atoms can remain or  react with other atoms to fill the up the valleys.
Thus, the diffusion-reaction mechanism automatically ``smooths"
the surface strongly at 15 K.   	

In addition to the rapid dropping in monolayer abundance of the four major carbon-bearing species, there is a second distinct feature at 15 K.  Both the monolayer abundances of  CO and H$_2$CO increase strongly towards the highest monolayers before suddenly dropping.  This effect occurs because it is relatively difficult to bury CO and H$_{2}$CO under higher layers due to their reactivity with OH radicals.


\subsection{Comparison with observations}
Our primary interest in this work has been to introduce a new method to combine rate equations and
microscopic Monte Carlo simulations in a
 gas-grain model of the chemistry of cold interstellar cores. Toward this goal,  a small grain-surface chemical network was chosen for simplicity. However, we
would still like to compare our results with astronomical observations.  Since, again for simplicity, we do not consider non-themal desorption mechanisms here, as we have done recently with our latest gas-grain rate-equation code \citep{rob}, it is  unnecessary to compare our gas-phase results with observations since the results are very similar to those of \cite{ruffle}.   Rather, it makes more sense at this stage to compare our mantle results with infrared absorption measurements.  These results are insensitive to non-thermal desorption, which only affects mantle abundances slightly. We are particularly interested in determining whether or not  a rough surface
model can help to obtain abundances closer to the observations.
In Table \ref{table7} we list calculated and observed fractional abundances for four ices: H$_{2}$O, CO$_{2}$, CO, and CH$_{3}$OH.  
The water abundance is tabulated as the fractional abundance with respect
to $n_{\rm H}$ while the abundances of the other ices are listed as percentages of the water ice abundance. Our theoretical
 results are calculated at $2 \times 10^5$ yr.  For comparison, master equation results at 10 K for a time of $ 3.2 \times 10^{5}$ yr are also listed \citep{herbst2}.  The observational data for W33 A are
from \citet{dartois} and \citet{gibb}, while the data for Elias 16 are from \citet{gibb}.

\begin{table*}
\parbox{12.0cm}{\caption{Calculated and observed mantle abundances$^{1}$}\label{table7}} \\
\begin{tabular}{llllllll}\hline \hline
Species  & W33A   & Elias 16 & Flat surface & Flat surface & Rough Surface & Rough surface & ME   \\ 
  &        &    &     10 K & 15 K  & 10 K & 15 K & 10 K \\ \hline   
H$_2$O   & 1.1(-4) & 1.25(-4)  & 1.8(-4)            & 1.8(-4)            & 1.5(-4)             & 2.1(-4) & 1.1(-4)\\
CO$_2$   & 13     & 18       & 4.5               & 10                & 7.6                & 6.2 & 2.6(-6) \\ 
CO       & 8      & 25       & 0.35              & 2.6               & 0.75               & 2.9 & 0.025 \\
CH$_3$OH & 18     & $<$3       & 11                & 1.7               & 9.5                & 3.9  & 35.7 \\ \hline \hline
\end{tabular}
\\
a(-b) refers to $\rm a\times 10^{-b}$\\
$^1$ water abundance with respect to $n_{\rm H}$; other ices listed as percentage of the water ice abundance\\
\end{table*}

As compared with the master equation results, which are based on a macroscopic stochastic approach in which no distinction is made for local site differences and differences according to monolayer, the Monte Carlo results are a distinct improvement.  The master equation abundance for CO  is very small at 10 K, although  better results are obtained at 20 K.  As for CO$_{2}$, the results are low at both 10 K and 20 K.  For the Monte Carlo models, agreement is reasonably good regardless of the type of surface model except perhaps for CO.  The agreement for CO$_{2}$ is especially important given the ubiquity and high abundance of this ice component.  The W33A results are closer to our 10 K model predictions whereas  the Elias 16 results are closer to the 15 K predictions.  The higher temperature leads to more CO and less methanol.  Still, CO is somewhat underproduced in the models.  For the rough surface, increasing the lateral bond strength
to 0.4$E_{\rm D}$ from its standard value of 0.1$E_{\rm D}$ can increase CO from 0.75 percent to 3.0 percent of the
H$_2$O ice while keeping other species in good agreement with observation at 10 K.  

\section{Conclusions}
For the first time, a gas-grain chemical simulation of a dense interstellar cloud core has been performed with a combined rate equation-microscopic Monte Carlo approach. In this approach, the local characteristics of the surface and all monolayers accreted onto it can be followed.  The
surface network is chosen to be the H, O and CO system, which has been widely studied and is the key network to form methanol on grain surfaces.   Our results for the total mantle abundances of the carbon-bearing species CO, CO$_{2}$, and CH$_{3}$OH as well as for water are in reasonable agreement with observation in two well-studied protostellar sources.  In fact, the agreement is superior to that achieved by use of a macroscopic Monte Carlo approach known as the master equation treatment \citep{herbst2}.  

The approach utilized here is based on a number of simplifications made possible mainly by the low temperatures (10 K, 15 K), the starting chemical abundances, and an unusual aspect of the gas-phase chemistry.  First, if we start with hydrogen mainly in molecular form, we need not consider subsequent H$_{2}$ formation since it is a small effect.  Secondly, at the low temperatures used here, the only reactive species that can evaporate from grain surfaces is atomic hydrogen.  Thirdly, the residual atomic hydrogen abundance in the gas does not strongly affect the abundances of all other gas-phase species.  These simplifications allow an iterative solution in which gas-phase abundances computed initially do not change significantly except for the case of atomic hydrogen.  Our iterative procedure involves starting with a gas-grain calculation in which no surface chemistry or H evaporation is allowed.  The time-dependent abundances of H, O, and CO are then allowed to deplete onto grain surfaces and the surface chemistry and evaporation of H are treated by the continuous-time random walk Monte Carlo procedure.  The evaporation rate of hydrogen is used to refine the abundance of atomic hydrogen in the gas by an average correction over the time period of the calculation, an approximation made possible by the neglect of H$_{2}$ formation.    This refined abundance is then used, if necessary, to redo the Monte Carlo calculation of the surface chemistry until convergence is achieved.  If the evaporation of atomic hydrogen from grain surfaces is strongly time dependent, a more complex bootstrap procedure would be called for.    Were other species to evaporate, as occurs at higher temperatures,  and were non-thermal desorption to be included, a process that occurs even at low temperatures, we would probably have to devise a more intricate, non-iterative, method of integrating the gas-phase abundances simultaneously with the Monte Carlo determination of surface abundances. Such improvements are under active consideration, including an approach similar to the effective rate coefficient method \citep{rob,chang2}.  Finally, only a small number of surface processes are currently used; inclusion of a much larger number of processes by the Monte Carlo procedure would be exceedingly time-consuming whether or not changes in gas-phase abundances are also calculated. Whether a suitable procedure that will allow us to remove all of these simplifications can be achieved in the near future remains uncertain at his point.

\begin{acknowledgements}
E. H. acknowledges the support
of the National Science Foundation (U. S.) for his research program
in astrochemistry.  We thank Robin Garrod and Donghui Quan for their assistance with the use of the Ohio State gas-grain code.
\end{acknowledgements}

{}


\begin{thebibliography}{}
\bibitem[{Al-Halabi et al.}(2002)]{alha}
Al-Halabi, A., Kleyn, A., van Dishoeck, E. F.,  \& Kroes, G. 2002 J. Phys. Chem. B, 106, 6515

\bibitem[{Awad et al.}(2005)]{awad}
Awad, Z., Chigai, T., Kimura, Y., Shalabiea, O. M., \& Yamamoto, T. 2005, ApJ. 626, 262  

\bibitem[{Biham et al.}(2001)]{biham}
Biham, O., Furman, I., Pirronello, V., \& Vidali, G. 2001, ApJ, 553, 595

\bibitem[{Buch \& Czerminski}(1991)]{buch}
Buch, V., \& Czerminski, R. 1991, J. Chem. Phys., 96, 6026

\bibitem[{Caselli et al.}(1998)]{caselli}
Caselli, P., Hasegawa, T. I. \& Herbst, E. 1998, ApJ, 495, 309

\bibitem[{Caselli et al.}(2002)]{caselli2}
Caselli, P., Stantcheva, T., Shalabiea, O., Shematovich, V. I., \& Herbst, E. 2002, P\&SS 50, 12

\bibitem[{Chang et al.}(2005)]{chang}
Chang, Q., Cuppen, H. M., \&  Herbst, E. 2005, A\& A, 434, 599

\bibitem[{Chang et al.}(2006)]{chang2}
Chang, Q., Cuppen, H. M., \&  Herbst, E. 2006, A\& A,  458, 497

\bibitem[{Charnley}(1998)]{charnley1}
Charnley, S. B. 1998, ApJ, 509, L121

\bibitem[{Charnley}(2001)]{charnley}
Charnley, S. B. 2001, ApJ, 562, L99

\bibitem[{Cuppen \& Herbst}(2005)]{cuppen}
Cuppen, H. M., \& Herbst, E. 2005, MNRAS, 361, 565

\bibitem[{Cuppen \& Herbst}(2007)]{cuppen2}
Cuppen, H. M., \& Herbst, E. 2007, ApJ,  submitted

\bibitem[{Dartois et al.}(1999)]{dartois}
Dartois, E., Schutte, W., Geballe, T.R. et al., 1999, A\&A, 342, L32

\bibitem[{Ehrenfreund \& Schutte}(2000)]{schutte}
Ehrenfreund, P., \& Schutte, W.A. 2000, in IAU Symp. 197,
Astrochemistry: From Molecular Cloud to Planerary systems, ed. Y.C. Minh \& E.F. van Dishoeck (San Francisco: ASP), 135

\bibitem[{Fuchs et al.}(2007)]{Fuchs:prep}
Fuchs, G. W., Ioppolo, S., Bisschop, S. E., van Dishoeck, E. F., \& Linnartz, H. 2007, to be submitted to A\&A.

\bibitem[{Garrod \& Herbst}(2006)]{garrod}
Garrod, R., \& Herbst, E. 2006, A\&A, 457, 927

\bibitem[{Garrod et al.}(2006)]{rob}
Garrod, R. T., Park, I.P., Caselli, P., \& Herbst, E. 2006, Faraday Discussions, 133, 51

\bibitem[{Gibb et al.}(2000)]{gibb}
Gibb E.L., Whittet, D.C.B., Schutte, W., et al. 2000, ApJ, 536, 347

\bibitem[{Green at al.}(2001)]{green}
Green, N. J. B., Toniazzo, T., Pilling, M. J., et al. 2001, A\&A, 375, 1111

\bibitem[{Hasegawa et al.}(1992)]{hasegawa}
Hasegawa, T.I., Herbst, E., \& Leung, C.M. 1992, ApJS, 82, 167

\bibitem[{Herbst}(1995)]{herbst1}
Herbst, E. 1995, Annu. Rev. Phys. Chem., 46, 27


\bibitem[{Hiraoka et al.}(2002)]{hiraoka}
Hiraoka, K., Sato, T., Sato, S., Sogoshi, N., Yokoyama, T., Takashima, H., \& Kitagawa, S. 2002, ApJ, 577, 265

\bibitem[{Hollenbach \& Salpeter}(1970)]{hollen}
Hollenbach, D., \& Salpeter, E. 1970, J. Chem. Phys., 53. 79

\bibitem[{Jansen}(1995)]{jansen}
Jansen, A.P.J. 1995, Comp. Phys. Comm., 86, 1

\bibitem[{Katz et al.}(1999)]{katz}
Katz, N., Furman, I., O., Pirronello, V., \& Vidali, G. 1999, ApJ, 522, 305

\bibitem[{Lohmar \& Krug}(2006)]{krug}
Lohmar, I., \&  Krug, J. 2006, MNRAS, 370, 1025

\bibitem[{Lipshtat \& Biham}(2004)]{biham2}
Lipshtat, A., \& Biham, O. 2004, PRL, 93, 170601


\bibitem[{Montroll \& Weiss}(1965)]{montroll}
Montroll, E., \& Weiss, G. H. 1965, J. Math. Phys., 6, 167

\bibitem[{Perets et al.}(2005)]{perets}
Perets, H. B., Biham, O., Manico, G., Pirronello, V. Roser, J., Swords, S., \& Vidali, G. 2005, ApJ, 627, 850

\bibitem[{Pickles \& Williams}(1977)]{pickles}
Pickles, J.B., \& Williams, D.A. 1977, Ap\&SS, 52, 443

\bibitem[{Roberts et al.}(2004)]{millar}
Roberts, H., Herbst, E., \& Millar, T. J. 2004, A\&A, 424, 905

\bibitem[{Ruffle \& Herbst}(2000)]{ruffle}
Ruffle, D. P., \& Herbst, E. 2000, MNRAS, 319, 837

\bibitem[{Stantcheva \& Herbst}(2004)]{herbst2}
Stantcheva, T., \& Herbst, E. 2004, A\&A, 421, 241

\bibitem[{Stantcheva et al.}(2002)]{herbst3}
Stantcheva, T., Shematovich, V. I. ,\& Herbst, E. 2002, A\&A, 391, 1069

\bibitem[{Tielens \& Hagen}(1982)]{tielen}
Tielens, A.G.G.M., \& Hagen W. 1982, A\&A, 114, 245


\bibitem[{Tielens \& Whittet}(1996)]{tielens}
Tielens, A.G.G.M., \& Whittet, D.C.B. 1996, IAU Symp. 178, Molecules in Astrophysics: probes and Processes, ed. E. F. van Dishoeck (Dordrecht:Kluwer), 45


\bibitem[{Wakelam et al.}(2006)]{wakelam}
Wakelam, V., Herbst, E., \& Selsis, F. 2006, A\&A, 451, 551

\bibitem[{Watanabe \& Kouchi}(2002)]{watanabe1}
Watanabe, N., \& Kouchi, A. 2002, ApJ, 571, L173

\bibitem[{Watanabe et al.}(2003)]{watanabe2}
Watanabe, N., Shiraki, T., \& Kouchi, A. 2003, ApJ, 588, L121
\end{thebibliography}
\end{document}